# Epitaxial growth and transport properties of a metallic altermagnet CrSb on a GaAs (001) substrate

Seiji Aota[1] and Masaaki Tanaka[1,2,3,*]

[1] *Department of Electrical Engineering and Information Systems, The University of Tokyo, 7-3-1 Hongo, Bunkyo-ku, Tokyo 113-8656, Japan*

[2] *Center for Spintronics Research Network (CSRN), The University of Tokyo, 7-3-1 Hongo, Bunkyo-ku, Tokyo 113-8656, Japan*

[3] *Institute for Nano Quantum Information Electronics, The University of Tokyo, 4-6-1 Komaba, Meguro-ku, Tokyo 153-0041, Japan*

[*] E-mail: masaaki@ee.t.u-tokyo.ac.jp

**Abstract**

A newly identified class of magnetic materials called altermagnets has attracted much attention due to the practical properties of spin-splitting bands akin to ferromagnets and small compensated magnetization akin to antiferromagnets. These features make them promising candidates for applications in spintronics devices. Among candidate materials, CrSb is promising because of its high ordering temperature (~705 K) and large spin-splitting energy; however, it is predicted that tuning the Néel vector requires additional symmetry breaking or a change in the easy magnetization axis. While applying epitaxial strain can modulate the symmetry, the selection of substrates with closely matched lattice constants for heteroepitaxial growth is limited for altermagnets, which generally have low crystal symmetry. Therefore, exploring the heteroepitaxial growth of altermagnet thin films on well-established, dissimilar crystal systems is valuable. (001)-oriented III-V semiconductors, which share group-V elements with the overgrown CrSb, offer an ideal platform because they are expected to have material compatibility with stable interfaces, as well as tunability of the buffer layer's bandgap and lattice constant by varying the atomic composition of their group-III and group-V atoms. In this study, we have achieved the molecular beam epitaxial growth of a single-crystalline CrSb $(\bar{1}10)$ thin film on a GaAs (001) substrate by inserting thin FeSb $(\bar{1}10)$ / AlAs (001) buffer layers. The in-plane epitaxial relationship is found to be CrSb [110] || GaAs [110] and CrSb [001] || GaAs $[1\bar{1}0]$, and epitaxial strain is also confirmed. We also characterized the magnetotransport properties of the grown CrSb thin film. Although the obtained conductivity tensors are mainly explained by a two-carrier model, not by an anomalous Hall effect, this model reveals the presence of high-mobility electron and hole carriers.

## I. INTRODUCTION

Antiferromagnets have long been considered intriguing but impractical due to the difficulty in detecting and manipulating their magnetic ordering. However, recent advances have demonstrated techniques for manipulating and detecting Néel vectors in several antiferromagnets [1–4]. These materials exhibit high-speed (terahertz order) spin dynamics [3–6] and compensated magnetization, both of which are expected to be useful for device applications. Recently, it has been shown that a new type of non-relativistic spin-splitting appears in some collinear antiferromagnets with a broken time-reversal ($T$) symmetry as well as broken $PT$ and $tT$ symmetries, where $P$ and $t$ are a space inversion operation and a lattice translation operation, respectively [7–9]. These symmetry breakings are only achieved by the presence of asymmetric non-magnetic elements and lead to spin-split bands akin to ferromagnets. Such materials are named "altermagnets" and are characterized by spin-splitting in momentum space, which gives rise to strong anomalous responses such as the spontaneous anomalous Hall effect (AHE) [7,10–12], spin-polarized current generation [13–18], and giant magnetoresistance and tunneling magnetoresistance [19,20]. These characteristics, combined with the advantages of antiferromagnets, show the potential for realizing innovative spintronics devices.

Among candidate materials, metallic $RuO_2$ [15–17,20,21] and semiconducting MnTe [11,12,22–25] have been actively studied. However, the anticipated magnetic ordering of $RuO_2$ has recently been questioned [26–28]. Meanwhile, for MnTe, spontaneous AHE and manipulation of its sign have been reported [11,12]. CrSb is also theoretically predicted to be a promising altermagnetic metal with a very large spin-splitting, and the validity of the theoretical calculations has been confirmed by angle-resolved photoemission spectroscopy (ARPES) in several reports [29-32]. CrSb is an A-type antiferromagnet with a NiAs-type crystal structure (the lattice constants: $a$ = 0.4103 nm, $c$ = 0.5463 nm [33]) like MnTe, and the antiparallel spin configuration along the $c$-axis is confirmed by neutron diffraction measurements [34]. The crystal structure, including the spin directions, is shown in Fig. 1(a). One can see that the state after $PT$ or $tT$ operation does not overlap with the original state due to the presence of the nonmagnetic Sb atoms. It is also worth noting that the magnetic ordering temperature is very high, ~705 K [34], and that the presence of Weyl points is predicted from band-structure calculations [32,35].

However, the detection and control of the Néel vector in altermagnets remains challenging. In altermagnets, the Néel vectors and the magnetic moment of each magnetic atom are aligned parallel to the easy axis of magnetization. For the easy axis, two directions of the Néel vector can be considered, with opposite spin arrangements, corresponding to each other by $T$ operation. The manipulation and detection of the Néel vector between these two states is critical for applications of altermagnets. In detecting Néel vectors, the AHE is an important phenomenon observed in some altermagnets. It has been recognized that there are multiple mechanisms to account for the AHE; in particular, an intrinsic origin related to the Berry curvature of occupied electronic bands, which does not necessarily require macroscopic magnetization, has been used to understand the appearance of the AHE in antiferromagnets with broken $T$ symmetry [10-12, 36]. Unfortunately not all altermagnets can show the AHE, and in some cases, its appearance is fundamentally prohibited by the magnetic crystal symmetry [10]. The AHE can be represented by the anomalous Hall vector $\boldsymbol{h}$, where the Hall current $\boldsymbol{j}_{\mathrm{Hall}}$ is given by $\boldsymbol{h} \times \boldsymbol{E}$, with $\boldsymbol{E}$ representing the electric field. As it plays a similar role to a magnetic field, $\boldsymbol{h}$ is a $T$-odd axial vector. For example, if the crystal has mirror planes, the sign of $\boldsymbol{h}$ components horizontal to the mirror planes are reversed by the mirror operation, so $\boldsymbol{h}$ must be perpendicular to the mirror plane. Thus, whether a crystal can have non-zero $\boldsymbol{h}$ or not depends on its crystal and spin symmetries. In fact, CrSb, which has an easy magnetization axis in the [001] direction ($c$-axis), cannot have the AHE in any direction due to the presence of three glide planes ($M \cdot t_{1/2}$ where $M$ is a mirror operation and $t_{1/2}$ is a half-unit cell translation) and three magnetic mirror planes ($M \cdot T$) orthogonal to them (see Fig. 1(a) and Fig. 2(a)). No $T$-odd axial vector in any direction satisfies all these magnetic mirror and glide symmetries [37-40]. This contrasts with MnTe, which has an easy magnetization axis in the $[\bar{1}10]$ direction [11,12]. Very recently, Zhou *et al.* [37] have demonstrated the sign reversal of the AHE in CrSb by breaking some of these symmetries by applying epitaxial strain. Controlling the easy magnetization axis, for example by appropriate doping, is another option. Although magnetic anisotropy is also changed by strain, axial strain parallel to the $a$-axis in the crystal is reported to be insufficient to change the easy axis [41].

It is also known that some antiferromagnets possess spontaneous spin canting magnetization (also called weak ferromagnetism), caused by the Dzyaloshinskii-Moriya interaction (DMI). If this weak ferromagnetism is coupled with the spin alignment of altermagnets, it is possible to manipulate the Néel vectors of the altermagnets by an external magnetic field. It should be noted, however, that the conditions under which weak ferromagnetism appears are not yet fully understood. In the formulated DMI ($H^{DM}{}_{i,j} = D_{ij} \cdot [S_i \times S_j]$, where $S_i$ and $S_j$ are neighboring spins), weak ferromagnetism does not appear in centrosymmetric crystals. Nevertheless, weak ferromagnetism has been reported in some centrosymmetric antiferromagnets, for example in MnTe. Several papers discussed its origins, including higher-order effects in DMI [42, 43]. In almost the same cases where anomalous transport in altermagnets is non-zero, the magnetic point group is compatible with weak ferromagnetism [43]. This can be understood from the fact that the weak ferromagnetic moment is also a $T$-odd axial vector as well as $\boldsymbol{h}$. In fact, unlike MnTe, no weak ferromagnetism has been observed in bulk CrSb [38-40], therefore the symmetry breaking by strain as in the case of inducing a non-zero AHE is essential to control the Néel vector of CrSb. In the presence of these AHE and weak ferromagnetism, the Néel vector with weak ferromagnetism is reversed between the two directions parallel to the CrSb [001] easy axis by sweeping an external magnetic field || $\boldsymbol{h}$. The reversal of the Néel vector is observed by the anomalous Hall resistance in a conventional Hall measurement [11,12,37] (See also Supplementary Fig. S7 [60] for illustration).

For this reason, symmetry change by applying epitaxial strain is important when developing the heteroepitaxial growth of altermagnetic CrSb thin films for practical devices. (001)-oriented III-V semiconductors are an ideal platform because they have good material compatibility (sharing group-V atoms), excellent tunability of the bandgap and the lattice constants by changing their atomic compositions [44], as well as easy integration with semiconductor devices. Single-crystal growth of CrSb (001) on a GaAs (111)A substrate [45] and CrSb $(\bar{1}10)$ on a GaAs (110) substrate [29] has been already reported, and the latter one can induce epitaxial strain which breaks the symmetries to enable the reversal of the Néel vector. However, there are significant advantages to growing CrSb on a GaAs (001) substrate: Easy control of the lattice constants with graded buffer layers [46,47], epitaxial lift-off techniques [48,49], and good compatibility and easy integration with III-V semiconductor electronics [50]. From the viewpoint of epitaxial growth of dissimilar materials, we refer to previous works reporting that MnAs, a transition metal pnictide with a NiAs crystal structure similar to CrSb, has been epitaxially grown on GaAs (001) substrates [51,52]. It has been shown that the growth direction of the MnAs crystal is significantly influenced by the first one or a few monolayers formed on the semiconductor surface. Single-crystal MnAs $(\bar{1}10)$ has been grown on GaAs by chemically modulating the GaAs surface with an As flux: The epitaxial relationship was found to be MnAs $(\bar{1}10)$ || GaAs (001) and || MnAs [001] || GaAs $[1\bar{1}0]$. In this study, we demonstrate the epitaxial growth of a CrSb thin film on GaAs (001) with the same epitaxial relationship by engineering buffer layers. The grown CrSb thin film is found to be single-crystalline with atomically smooth and abrupt interfaces, and there is an axial compressive strain along the CrSb [110] direction. Since a detailed study of the transport properties in a single crystalline CrSb thin film has been lacking, we also investigate the magnetotransport properties of the epitaxial CrSb thin film grown on GaAs (001).

## II. METHODS

### A. Sample preparation and characterization

We grew CrSb thin films via molecular beam epitaxy (MBE), where Ga, Al, Cr, and Fe fluxes were supplied from Knudsen effusion cells, and As and Sb fluxes were supplied from valved cracking cells. During growth, the cracking temperatures of the As and Sb cells were set at 600°C and 900°C, respectively, providing $As_4$ and $Sb_4$ fluxes. The crystal structure of CrSb, epitaxial relationship with GaAs (001), and sample structure are shown in Figs. 1(a)–1(c). After removing the surface native oxide at 580°C, a 50-nm-thick GaAs buffer layer was first grown on a semi-insulating GaAs (001) substrate at 550°C to obtain an atomically smooth surface. Next, a 5-nm-thick AlAs layer was grown on the GaAs buffer layer with the same substrate temperature. Then, one monolayer (ML)-thick FeSb was grown on the AlAs layer at 270 °C. After that, a 5-nm-thick CrSb layer was grown at 270°C, followed by the growth of a 25-nm-thick CrSb layer at 340°C, resulting in a total CrSb thickness of 30 nm. We used this two-step growth procedure; at the initial stage of the growth, we grew a very thin 5-nm-thick CrSb layer at lower temperature 270°C to obtain smooth surface preventing island growth, then another 25-nm-thick CrSb layer at higher growth temperature 340°C to improve the surface smoothness and film quality. We did not observe any interface between the CrSb layers grown at two different temperatures, indicating that uniform CrSb is obtained (see Supplementary Fig. S2 [60]). We found that insertion of 1 ML FeSb effectively stabilizes the $(\bar{1}10)$ growth plane of CrSb. During growth of CrSb (or FeSb), the atomic flux of Sb was adjusted to be 1:1 with Cr (or Fe). After the MBE growth, the samples were annealed *in-situ* at 400°C for 15 minutes to improve the crystal quality. During the growth, the surface conditions were monitored by reflection high energy electron diffraction (RHEED). The crystal structures of the samples were characterized by x-ray diffraction (XRD) with a Cu Kα ($\lambda = 0.1542$ nm) radiation source. Reciprocal space mappings (RSM) around the $(1\bar{2}0)$ and the $(\bar{3}34)$ peaks of CrSb were used to estimate the lattice constants. We also characterized the sample by high-resolution lattice images using scanning transmission electron microscopy (STEM). Magnetization and transport properties were measured by a superconducting quantum interference device (SQUID) and physical property measurement system (PPMS), respectively.

**B. Calculation details**

We used first-principles calculations based on the density-functional theory (DFT) to study the band structure and transport properties of CrSb. We used the Quantum Espresso code [53-55] and GGA+U (we used $U$ =0.25 eV following a previous calculation that reproduced experimental lattice constants and magnetic moments [33,41,55]), and the projector-augmented wave potentials (PAW). The calculations were carried out using the lattice constants estimated in our experiment, with only the Sb atoms in the crystal subject to structural optimization. For the band structure calculation, the electron wave function was expanded in plane waves up to a cut-off energy of 60 Ry (816 eV), and a 9×9×9 k-point grid was used for the Brillouin-zone integration. By using maximally localized Wannier functions [56,57], we have constructed an effective tight-binding Hamiltonian to calculate the Fermi surface and the anomalous Hall conductivity (AHC) tensor. Integration of the Berry curvature was carried out within a 512×512×512 *k*-point grid with the Wannier Berri code [58]. For mean absolute error (MAE) calculations, a *k*-point grid of 16×16×16 was used for the accuracy and the force theorem method [59] was used for calculating the total energy difference by changing the Néel vector direction. The norm-conserved (NC) potentials with higher cut-off energy of 100 Ry (136 eV) were used for the MAE calculations instead of the PAW potentials due to the software compatibility

## III. Results and discussion

**A. MBE growth and structural characterizations**

The target growth orientation of the CrSb thin film is shown in Fig. 1(b). Against our expectation, we were not able to obtain a single-crystalline CrSb $(\bar{1}10)$ thin film when we tried to grow CrSb directly on a GaAs (001) or on a GaSb (001) substrate regardless of the surface modulation with an As (or Sb) flux. However, when CrSb was grown on an AlAs buffer layer on a GaAs (001) substrate, a $(\bar{1}10)$ oriented CrSb thin film was grown, and CrSb($\bar{h}h0$) peaks were observed. Nevertheless, additional peaks corresponding to $(\bar{1}11)$ and $(\bar{1}12)$ were also detected in the XRD spectra, as shown in Supplementary Fig. S1 [60]. To overcome this problem, we inserted a 1-ML-thick FeSb layer beneath the CrSb layer. FeSb has the same NiAs-type crystal structure and lattice constants ($a$ = 0.4066 nm, $c$ = 0.5133 nm [61]), which are close to those of CrSb. By incorporating this FeSb layer, we successfully achieved the growth of a single-crystalline CrSb $(\bar{1}10)$ thin film. This final sample structure is (from top to bottom) CrSb (30 nm) / FeSb (1 ML) / AlAs (5 nm) / GaAs(001) substrate, as illustrated in Fig. 1(c). We also confirmed that a 1-ML-thick AlAs and 1-ML-thick FeSb layers are necessary for stabilizing the targeting single-crystal growth of CrSb $(\bar{1}10)$. This finding is consistent with the previous studies showing that the crystal orientation of a metallic thin film grown on a structurally dissimilar semiconductor substrate critically depends on the first few MLs of growth. Growing FeSb or other NiAs-type materials as a template with a thickness of one or a few MLs may be generally applicable for growing CrSb / III-V semiconductor heterostructures with desired orientations. The schematic sample structure and the RHEED patterns after the growth are shown in Fig. 1(c). Figure 1(d) shows a high-resolution STEM lattice image of the grown sample, indicating the epitaxial relationship of CrSb $(\bar{1}10)$ || GaAs (001), CrSb $[00\bar{1}]$ || GaAs $[\bar{1}10]$, and CrSb [110] || GaAs [110], as intended. The XRD result in Fig. 1(e) shows only diffraction peaks of CrSb $(\bar{h}h0)$ and the GaAs (001) substrate. The sharp peaks of CrSb $(\bar{1}10)$ further support the single crystallinity of the grown CrSb film. Figure 1(d) shows a clear lattice image of CrSb expected from the NiAs-type crystal structure. Interestingly, despite the large lattice mismatch between the $c$-axis of CrSb (0.5463 nm [34], see vertical red lines in Fig. 1(d)) and the $a/\sqrt{2}$ of AlAs/GaAs (0.3997 nm, see green vertical lines in Fig. 1(d)), an atomically abrupt and flat interface was obtained; this means that 7 layers of the GaAs $(\bar{1}10)$ plane correspond to 5 layers of the CrSb (001) plane. The lattice constants of CrSb obtained from the RSM measurement are $2d_{110}$ = 0.4036 nm, $d_{\bar{1}10}$ = 0.3530 nm and $d_{001}$ = 0.5634 nm, where $d_{110}$, $d_{\bar{1}10}$, and $d_{001}$ are the lattice spacings in the direction of the Miller indices shown in the subscript. This indicates that the CrSb crystal is mainly compressed along the [110] direction and elongated in the [001] direction compared to the bulk lattice constants ($a = 2d_{110}$= 0.4103 nm and $c = d_{001}$= 0.5463 nm [34]).

Figure 2(a) shows schematic crystal structures of the CrSb film projected along the [001] direction (side views, left panel: unstrained, center panel: strained) and along the [110] direction (side view, right panel: strained), where symmetry planes of CrSb are shown by cyan, magenta, and yellow lines. Red dots and arrows denote the magnetic moments of Cr atoms. The strain applied along the [110] direction breaks some of the magnetic mirror planes and the glide planes as represented by the four dashed lines (cyan and magenta), as shown in the center figure of Fig. 2(a). Since a mirror operation ($M$) and a glide mirror operation ($M \cdot t_{1/2}$) reverse $T$-odd axial vectors parallel to them, the symmetry breaking allows the appearance of non-zero $T$-odd axial vectors, such as anomalous Hall vector and weak ferromagnetic moment parallel to the CrSb $[\bar{1}10]$ direction (see the orange double-headed arrows). We define the strain $\varepsilon$ in this paper as

$$\varepsilon = \left(\frac{\sqrt{3}d_{110}}{d_{\bar{1}10}} - 1\right) \times 100\ (\%). \quad (1)$$

Here, $\varepsilon$ is defined as the rate of change in $d_{110}$ compared to the length calculated form the height ($d_{\bar{1}10}$) assuming a regular hexagon (see Supplementary Fig. S6(b) [60]). This is because the size of the hexagonal plane is also affected by the strain in the [001] direction. In our definition, $\varepsilon$ is estimated to be $-1.0\%$ for our CrSb film.

### B. Band structure calculation

We performed theoretical calculations to examine the influence of the strain on the electronic state and the magnitude of the AHC. The crystal is no longer hexagonal and can be regarded as base-center orthorhombic. Band structure calculations were carried out along the corresponding common high-symmetry paths of the original hexagonal crystal and the additional T–Γ–T' path, as shown in Fig. 2(b). The high-symmetry points are labeled in Fig. 2(c). The spin-up and spin-down bands in the high-symmetry path are almost or completely degenerated (see purple curves in Fig. 2(b)), while a spin-polarized band splitting is clearly observed between the Γ point and the T (T') point (see red and blue curves in Fig. 2(b)). Figure 2(c) shows the Fermi surfaces for holes and electrons in $k$-space. In comparison with a previous study [39] and calculations shown in Supplementary Fig. S4 [60], which were obtained using bulk lattice parameters, the band structure is not significantly changed but the electron Fermi surface expands and forms honeycomb-shaped structures, as shown in Fig. 2(c). The calculated AHC as a function of the Fermi energy ($E_F$) is shown in Fig. 2(d). The AHC value ($\sigma_{yx}^{\mathrm{AHE}}$) is estimated to be about 20 S/cm around $E_F = 0$. Note that we use a notation where $x$ corresponds to the [001] axis, $y$ to the [110] axis, and $z$ to the $[\bar{1}10]$ axis of CrSb.

### C. Magnetic properties

We also calculated the magnetic anisotropy energy of CrSb using the obtained lattice parameters in our thin film and those of the bulk, as listed in Table. I. The change in magnetic anisotropy with strain is small, and the easy magnetization axis of the thin film is not expected to change from the [001] direction in our calculation. Magnetization ($M$ – $H$) curves were measured as shown in Fig. 3. The diamagnetic contribution of the substrate was subtracted from the raw data (see Supplementary Fig. S5 [60] for the procedure and raw data before the subtraction). We found a tiny ferromagnetic component of ~10 emu/cc (= $4.3 \times 10^{-2}$ $\mu_B$/Cr atom), but this is much larger than the values reported for weak ferromagnetism in MnTe (~5 $\times 10^{-5}$ $\mu_B$/Mn atom) [12]. It is very difficult to determine whether this is weak ferromagnetism or not, since CrSb has a high Néel temperature over 700 K. But at least this value $4.3 \times 10^{-2}$ $\mu_B$/Cr atom in our sample is much smaller than the magnetization that would be expected if CrSb had ferromagnetic spin ordering (=2.7 $\mu_B$/Cr atom [35]), and thus this possibility is ruled out. No response corresponding to this ferromagnetic component has been observed in the subsequent magnetotransport measurement, suggesting that we may have observed magnetization of localized areas which does not contribute to the transport properties. Instead of weak ferromagnetism, surface magnetization may be a possible origin, as reported in $Cr_2O_3$ [62], for example.

**D. Magnetotransport properties**

We measured the temperature dependence of the longitudinal resistivity $\rho_{xx}$ of the grown CrSb film, as shown in Fig. 4(a). The residual resistivity ratio (RRR = $\rho_{300K}$ / $\rho_{5K}$) is 4.30. Although the RRR is smaller than that reported in bulk CrSb (~7 in [39] and 8.94 in [40]), it is higher than the RRR around 1 reported by Zhou *et al.* [37], and our film shows high conductivity and temperature dependence similar to that of bulk CrSb. The magnetoresistance (MR) and the Hall resistivity $\rho_{yx}$ were also measured at various temperatures ranging from 5 K to 300 K. The positive MR reached 1.4 % at 5 K and 7 T, as shown in Fig. 4(b). The slope of $\rho_{yx}$ is changed from positive at room temperature to negative at low temperature, as shown in Fig. 4(c), indicating the coexistence of holes and electrons. Fig. 4(d) shows the AHE component obtained by subtracting the ordinary Hall effect (OHE) contribution calculated using the carrier density and mobility by the multicarrier analysis (as described below). The expected AHE vector is perpendicular to the sample, but no hysteresis of the Hall resistivity was observed. To further investigate the origin of the Hall resistivity characteristics, we carried out multiple carrier fittings [38-40] to the $\rho_{xx}$ and $\rho_{yx}$ data. The longitudinal conductivity $\sigma_{xx}$ and Hall conductivity $\sigma_{yx}$ are expressed by $\sigma_{xx} = \frac{\rho_{xx}}{\rho_{xx}^2+\rho_{yx}^2}$ and $\sigma_{yx} = \frac{-\rho_{yx}}{\rho_{xx}^2+\rho_{yx}^2}$, respectively. Note that we assumed isotropic resistivity ($\rho_{xx} = \rho_{yy}$) and Onsager's relations ($\rho_{xy}(B) = -\rho_{yx}(B)$), which is not strictly true for $T$-breaking CrSb when the contributions to conductivity from the opposite Néel vectors are not equal. This model incorporates multiple types of carriers, characterized by carrier densities and mobilities, using the following equations:

$$\sigma_{xx}(H) = \Sigma_i \frac{n_i e \mu_i}{1 + \mu_i^2 (\mu_0 H)^2}, \quad (2)$$

$$\sigma_{yx}(H) = \Sigma_i \frac{s_n n_i e {\mu_i}^2 \mu_0 H}{1 + \mu_i^2 (\mu_0 H)^2}, \quad (3)$$

where $n_i$ and $\mu_i$ represent the carrier concentration and mobility, respectively, and $s_n$ takes $-1$ for electron-type carriers and +1 for hole-type carriers.

Figures 5(a) and 5(b) show the two-carrier fitting result for the conductivities $\sigma_{xx}$ and $\sigma_{yx}$, respectively, measured at 5 K and 300 K. At 300 K, slight deviation in $\sigma_{xx}$ in the low-field region was observed, but at other temperatures, both $\sigma_{xx}$ and $\sigma_{yx}$ can be well fitted using the two-carrier (electron and hole) model. The carrier density and mobility obtained from the two-carrier fittings for all temperatures are shown in Figs. 5(c) and 5(d). Although high mobility carriers above $10^3$ cm$^2$/Vs [38-40] reported in bulk CrSb have not been observed, our CrSb thin film also shows semi-metallic properties with both types of carriers, an electron mobility of $1.8 \times 10^2$ cm$^2$/Vs and a hole mobility of $1.6 \times 10^2$ cm$^2$/Vs, and carrier densities 4.5 ~ 6 x $10^{20}$ /cm$^3$ which are lower than those of typical metals. The small nonlinearity of the Hall resistivity compared with that of bulk CrSb [38-40] is considered to be due to the lower mobilities.

As can be seen from the analysis of the transport characteristics, the Hall resistivity can be explained by the two-carrier model, and it is inferred that the spontaneous AHE is not observed. The AHC values obtained from the theoretical calculation (see Fig. 2(d)) are sufficiently large to be detected within the measurement accuracy of this work. A possible reason for the absence of spontaneous AHE in our sample is that the reversal of the Néel vector has not been successful. Kluczyk *et al.* [12] proposed a model for the reversal of the Néel vector in MnTe, considering the incorporation of exchange field, anisotropic field, and DMI-like field, and argued that when the DMI-like field is too small in comparison with the exchange field, the spin cannot be flipped similarly to the case of a fully compensated antiferromagnet. In the case of CrSb, we expect that stain induces the DMI-like field, but it is unknown how much strain is required. The difference between our CrSb thin film and the sample corresponding to "Configuration II" of the recent study by Zhou *et al.* [37] is that the strain is opposite (–1.0% in our sample, +1.0% in ref. [37], using the same definition of Eq. (1)). It is difficult to understand how this difference affected the results, because the mechanism of the DMI-like field (weak ferromagnetism) is not yet clear. Further control of the strain by engineering the III-V semiconductor buffer layers utilizing the advantage of the MBE growth on a GaAs (001) substrate is necessary in our future work. The crystallinity is also an important factor. Although the magnetic configuration is different, some *T*-broken non-collinear antiferromagnets are known to show AHE, and it has been reported that the presence or absence of AHE strongly depends on the annealing temperature [63]. We have grown another sample and then annealed it at higher temperature of 550°C while supplying a Sb flux of $1.0 \times 10^{-4}$ Pa to prevent re-evaporation of Sb atoms, but no spontaneous AHE was observed as shown in Supplementary Fig. S6. It is difficult to increase further the annealing temperature because a higher annealing temperature would decrease the strain and lead to the appearance of a secondary phase (FeAs [64,65] and CrAs [66] or their mixed crystals are possible secondary phase candidates. See Supplementary Fig. S6(a) for the XRD data.), but there is still room for adjustment of the substrate temperature and Sb flux during the growth of CrSb, and selection of the III-V semiconductor buffer layer.

## IV. CONCLUSION

We have successfully grown a single-crystalline CrSb $(\bar{1}10)$ thin film on a GaAs (001) substrate, revealing its high crystal quality, atomically smooth and abrupt interface, and compressive strain of $\varepsilon = -1.0\%$ in the CrSb [110] direction. The longitudinal and Hall conductivities obtained in the magneto-transport measurements are explained by a two-carrier model, but a spontaneous AHE expected from the theoretical calculation with obtained lattice constants was not observed. Further increase of strain may be required to achieve a reversal of the Néel vector. The magnetic-field dependence of the Hall resistivity and the transition from p-type at room temperature to n-type at low temperature are explained by the competing contributions of holes and electrons. The obtained carrier concentrations and mobilities suggest that the grown CrSb film behaves like a semimetal with a maximum mobility of $1.8 \times 10^2$ $cm^2/Vs$. These results are consistent with previous studies of bulk CrSb and support that our CrSb thin film has the same altermagnetic ordering. Our study of a new growth method will contribute to further strain control for realizing useful functionalities and device applications of CrSb / III-V heterostructures.

## ACKNOWLEDGEMENTS

The authors thank Prof. Ryosho Nakane for his valuable advice on the epitaxial growth of CrSb. The authors also thank Prof. Le Duc Anh and Prof. Shinobu Ohya for their discussion and help. This work was partly supported by World-Leading Innovative Graduate Study Program for Quantum and Semiconductor Science & Technology Program (WINGS-QSTEP), Grants-in-Aid for Scientific Research (Grants Nos. 20H05650, 23K17324, and 25H00840), CREST Program (JPMJCR1777) of JST, and Spintronics Research Network of Japan (Spin-RNJ).

## REFERENCES

[1] T. Jungwirth, X. Marti, P. Wadley, and J. Wunderlich, Antiferromagnetic spintronics, Nat. Nanotech. **11**, 231 (2016).
[2] P. Wadley *et al.*, Electrical switching of an antiferromagnet, Science **351**, 587 (2016).
[3] V. Baltz, A. Manchon, M. Tsoi, T. Moriyama, T. Ono, and Y. Tserkovnyak, Antiferromagnetic spintronics, Rev. Mod. Phys. **90**, 15005 (2018).
[4] K. Olejník et al., Terahertz electrical writing speed in an antiferromagnetic memory, Sci. Adv. **4**, eaar3566 (2018).

[5] M. Fiebig, N. P. Duong, T. Satoh, B. B. Van Aken, K. Miyano, Y. Tomioka, and Y. Tokura, Ultrafast magnetization dynamics of antiferromagnetic compounds, J. Phys. D : Appl. Phys. **41**, 164005 (2008).
[6] H. Qiu *et al.*, Ultrafast spin current generated from an antiferromagnet, Nat. Phys. **17**, 388 (2021).
[7] L. Šmejkal, R. González-Hernández, T. Jungwirth, and J. Sinova, Crystal time-reversal symmetry breaking and spontaneous Hall effect in collinear antiferromagnets, Sci. Adv. **6**, eaaz8809 (2020).
[8] S. Hayami, Y. Yanagi, and H. Kusunose, Momentum-dependent spin splitting by collinear antiferromagnetic ordering, J. Phys. Soc. Jpn. **88**, 123702 (2019).
[9] LD. Yuan, Z. Wang, JW. Luo, E. I. Rashba, and A. Zunger, Giant momentum-dependent spin splitting in centrosymmetric low-Z antiferromagnets, Phys. Rev. B **102**, 14422 (2020).
[10] L. Šmejkal, A. H. MacDonald, J. Sinova, S. Nakatsuji, and T. Jungwirth, Anomalous hall antiferromagnets, Nat. Rev. Mater. **7**, 482 (2022).
[11] R. Gonzalez Betancourt *et al.*, Spontaneous anomalous Hall effect arising from an unconventional compensated magnetic phase in a semiconductor, Phys. Rev. Lett. **130**, 36702 (2023).
[12] K. P. Kluczyk *et al.*, Coexistence of anomalous Hall effect and weak magnetization in a nominally collinear antiferromagnet MnTe, Phys. Rev. B **110**, 155201 (2024).
[13] R. González-Hernández, L. Šmejkal, K. Vỳbornỳ, Y. Yahagi, J. Sinova, T. Jungwirth, and J. Železnỳ, Efficient electrical spin splitter based on nonrelativistic collinear antiferromagnetism, Phys. Rev. Lett. **126**, 127701 (2021).
[14] E. W. Hodt and J. Linder, Spin pumping in an altermagnet/normal-metal bilayer, Phys. Rev. B **109**, 174438 (2024).
[15] A. Bose *et al.*, Tilted spin current generated by the collinear antiferromagnet ruthenium dioxide, Nat. Electron. **5**, 267 (2022).
[16] H. Bai *et al.*, Observation of spin splitting torque in a collinear antiferromagnet $RuO_2$, Phys. Rev. Lett. **128**, 197202 (2022).
[17] S. Karube, T. Tanaka, D. Sugawara, N. Kadoguchi, M. Kohda, and J. Nitta, Observation of spin-splitter torque in collinear antiferromagnetic $RuO_2$, Phys. Rev. Lett. **129**, 137201 (2022).
[18] HY. Ma, M. Hu, N. Li, J. Liu, W. Yao, JF. Jia, and J. Liu, Multifunctional antiferromagnetic materials with giant piezomagnetism and noncollinear spin current, Nat. Commun. **12**, 2846 (2021).
[19] L. Šmejkal, A. B. Hellenes, R. González-Hernández, J. Sinova, and T. Jungwirth, Giant and tunneling magnetoresistance in unconventional collinear antiferromagnets with nonrelativistic spin-momentum coupling, Phys. Rev. X **12**, 11028 (2022).
[20] YY. Jiang, ZA. Wang, K. Samanta, SH. Zhang, RC. Xiao, W. Lu, Y. Sun, E. Y. Tsymbal, and DF. Shao, Prediction of giant tunneling magnetoresistance in $RuO_2/TiO_2/RuO_2$ (110) antiferromagnetic tunnel junctions, Phys. Rev. B **108**, 174439 (2023).
[21] O. Fedchenko *et al.*, Observation of time-reversal symmetry breaking in the band structure of altermagnetic $RuO_2$, Sci. Adv. **10**, eadj4883 (2024).
[22] J. Krempaskỳ *et al.*, Altermagnetic lifting of Kramers spin degeneracy, Nature **626**, 517 (2024).
[23] S. Lee *et al.*, Broken kramers degeneracy in altermagnetic MnTe, Phys. Rev. Lett. **132**, 36702 (2024).
[24] T. Osumi, S. Souma, T. Aoyama, K. Yamauchi, A. Honma, K. Nakayama, T. Takahashi, K. Ohgushi, and T. Sato, Observation of a giant band splitting in altermagnetic MnTe, Phys. Rev. B **109**, 115102 (2024).
[25] T. Aoyama and K. Ohgushi, Piezomagnetic properties in altermagnetic MnTe, Phys. Rev. Mater. **8**, L41402 (2024).
[26] S. Lovesey, D. Khalyavin, and G. van der Laan, Magnetic structure of $RuO_2$ in view of altermagnetism, Phys. Rev. B **108**, L121103 (2023).
[27] M. Hiraishi, H. Okabe, A. Koda, R. Kadono, T. Muroi, D. Hirai, and Z. Hiroi, Nonmagnetic Ground State in $RuO_2$ Revealed by Muon Spin Rotation, Phys. Rev. Lett. **132**, 166702 (2024).
[28] P. Keßler *et al.*, Absence of magnetic order in $RuO_2$: insights from $\mu$SR spectroscopy and neutron diffraction, npj Spintronics **2**, 50 (2024).
[29] S. Reimers *et al.*, Direct observation of altermagnetic band splitting in CrSb thin films, Nat. Commun. **15**, 2116 (2024).

[30] J. Ding *et al.*, Large Band Splitting in g-Wave Altermagnet CrSb, Phys. Rev. Lett. **133**, 206401 (2024).
[31] G. Yang *et al.*, Three-dimensional mapping of the altermagnetic spin splitting in CrSb, Nat. Commun. **16**, 1442 (2025).
[32] W. Lu *et al.*, Signature of topological surface bands in altermag- netic Weyl semimetal CrSb, Nano Lett. **25**, 7343 (2024).
[33] A. Kallel, H. Boller, and E. Bertaut, Helimagnetism in MnP-type compounds: MnP, FeP, CrAs and $CrAs_{1-x}Sb_x$ mixed crystals, J. Phys. Chem. Solids **35**, 1139 (1974).
[34] T. Hirone, S. Maeda, I. Tsubokawa, and N. Tsuya, On the magnetic properties of the system MnSb–CrSb, J. Phys. Soc. Jpn. **11**, 1083 (1956).
[35] J. Wu, Z. Zhang, J. Liu, and X. Shao, Magnetic quadratic nodal line with spin–orbital coupling in CrSb, Appl. Phys. Lett. **123**, 052407 (2023).
[36] S. Nakatsuji, N. Kiyohara, and T. Higo, Large anomalous Hall effect in a non-collinear antiferromagnet at room temperature. Nature **527**, 212–215 (2015).
[37] Z. Zhou, X. Cheng, M. Hu, J. Liu, F. Pan, and C. Song, Manipulation of the altermagnetic order in CrSb via crystal symmetry. Nature **638**, 645–650 (2025).
[38] T. Urata, W. Hattori, and H. Ikuta, High mobility charge transport in a multicarrier altermagnet CrSb, Phys. Rev. Mater. **8**, 84412 (2024).
[39] Y. Bai, X. Xiang, S. Pan, S. Zhang, H. Chen, X. Chen, Z. Han, G. Xu, and F. Xu, Nonlinear field dependence of Hall effect and high-mobility multi-carrier transport in an altermagnet CrSb, Appl. Phys. Lett. **126**, 042402 (2025).
[40] X. Peng et al., Scaling behavior of magnetoresistance and Hall resistivity in the altermagnet CrSb, Phys. Rev. B **111**, 144402 (2025)
[41] I. J. Park, S. Kwon, and R. K. Lake, Effects of filling, strain, and electric field on the Néel vector in antiferromagnetic CrSb, Phys. Rev. B **102**, 224426 (2020).
[42] C. Autieri, R. M. Sattigeri, G. Cuono, and A. Fakhredine. "Staggered Dzyaloshinskii-Moriya interaction inducing weak ferromagnetism in centrosymmetric altermagnets and weak ferrimagnetism in noncentrosymmetric altermagnets"
[43] I. I. Mazin, K. D. Belashchenko, "Origin of the gossamer ferromagnetism in MnTe Phys. Rev. B **110**, 214436 (2024).
[44] T. Glisson, J. Hauser, M. Littlejohn, and C. Williams, Energy bandgap and lattice constant contours of III–V quaternary alloys, J. Electron. Mater. **7**, 1 (1978).
[45] C. W. Burrows, J. D. Aldous, and G. R. Bell, Epitaxial growth and surface reconstruction of CrSb (0001), Results Phys. **12**, 1783 (2019).
[46] H. Choi, Y. Jeong, J. Cho, and M. Jeon, Effectiveness of non-linear graded buffers for In (Ga, Al) As metamorphic layers grown on GaAs (0 0 1), J. Cryst. Growth **311**, 1091 (2009).
[47] K. E. Lee and E. A. Fitzgerald, High-quality metamorphic compositionally graded InGaAs buffers, J. Cryst. Growth **312**, 250 (2010).
[48] K. Lee, J. D. Zimmerman, X. Xiao, K. Sun, and S. R. Forrest, Reuse of GaAs substrates for epitaxial lift-off by employing protection layers, J. Appl. Phys. **111**, 033527(2012).
[49] C.-W. Cheng, K.-T. Shiu, N. Li, S.-J. Han, L. Shi, and D. K. Sadana, Epitaxial lift-off process for gallium arsenide substrate reuse and flexible electronics, Nat. Commun. **4**, 1577 (2013).
[50] J. A. Del Alamo, Nanometre-scale electronics with III–V compound semiconductors, Nature **479**, 317 (2011).
[51] M. Tanaka, J. Harbison, T. Sands, T. Cheeks, V. Keramidas, and G. Rothberg, Molecular beam epitaxy of MnAs thin films on GaAs, J. Vac. Sci. Technol., B: Microelectron. Nanometer Struct. Process. Meas. Phenom. **12**, 1091 (1994).
[52] M. Tanaka, Ferromagnet (MnAs)/III–V semiconductor hybrid structures, Semicond. Sci. Technol. **17**, 327 (2002).
[53] P. Giannozzi *et al.*, QUANTUM ESPRESSO: a modular and open-source software project for quantum simulations of materials, J. Condens. Matter Phys. **21**, 395502 (2009).
[54] P. Giannozzi *et al.*, Advanced capabilities for materials modelling with QUANTUM ESPRESSO, J. Condens. Matter Phys. **29**, 465901 (2017).

[55] A. I. Snow, Neutron Diffraction Investigation of the Atomic Magnetic Moment Orientation in the Antiferromagnetic Compound CrSb, Phys. Rev. **85**, 365 (1952).
[56] A. A. Mostofi, J. R. Yates, YS. Lee, I. Souza, D. Vanderbilt, and N. Marzari, wannier90: A tool for obtaining maximally-localised Wannier functions, Comput. Phys. Commun. **178**, 685 (2008).
[57] G. Pizzi *et al.*, Wannier90 as a community code: new features and applications, J. Condens. Matter Phys. **32**, 165902 (2020).
[58] S. S. Tsirkin, High performance Wannier interpolation of Berry curvature and related quantities with WannierBerri code, npj Comput. Mater. **7**, 33 (2021).
[59] D. Li, C. Barreteau, M. R. Castell, F. Silly, and A. Smogunov, Out-versus in-plane magnetic anisotropy of free Fe and Co nanocrystals: Tight-binding and first-principles studies, Phys. Rev. B **90**, 205409 (2014).
[60] See Supplementary Material at [URL link] for additional information. A: Effect of 1 monolayer FeSb insertion, B: Additional structural characterization data, C: Band structure calculation of CrSb without strain, D: Measurement method of the magnetization (*M-H*) curve of the CrSb layer, E: CrSb with higher annealing temperature, F: Reversal of the Néel vector in altermagnets.
[61] K. Komędera, A. Jasek, A. Błachowski, K. Ruebenbauer, and A. Krztoń-Maziopa, Magnetic anisotropy in FeSb studied by $^{57}$Fe Mössbauer spectroscopy, J. Magn. Magn. Mater. **399**, 221 (2016).
[62] X. He, Y. Wang, N. Wu, A. N. Caruso, E. Vescovo, K. D. Belashchenko, P. A. Dowben, and C.Binek, Robust isothermal electric control of exchange bias at room temperature, Nat. mater. **9**, 579 (2010).
[63] H. Iwaki, M. Kimata, T. Ikebuchi, Y. Kobayashi, K. Oda, Y. Shiota, T. Ono, and T. Moriyama, Large anomalous Hall effect in $L1_2$-ordered antiferromagnetic $Mn_3Ir$ thin films, Appl. Phys. Lett. **116** (2020).
[64] K. Selte, A. Kjekshus, J. Sletten, and others, Crystal structure of FeAs, Acta Chem. Scand. **23**, 2047 (1969).
[65] A. Jamil, H. Noguchi, and H. Munekata, Growth of Fe–As crystalline films on GaAs (100) by molecular beam epitaxy, Thin Solid Films **516**, 3015 (2008).
[66] Z. Yu, W. Wu, Q. Hu, J. Zhao, C. Li, K. Yang, J. Cheng, J. Luo, L. Wang, and HK. Mao, Anomalous anisotropic compression behavior of superconducting CrAs under high pressure, PNAS **112**, 14766 (2015).
[67] K. Momma and F. Izumi, VESTA 3 for three-dimensional visualization of crystal, volumetric and morphology data, J. Appl. Crystallogr. **44**, 1272 (2011).
[68] M. V. Lebedev, E. Mankel, T. Mayer, and W. Jaegermann, Wet etching of GaAs (100) in acidic and basic solutions: a synchrotron- photoemission spectroscopy study, J. Phys. Chem. C **112**, 18510 (2008).

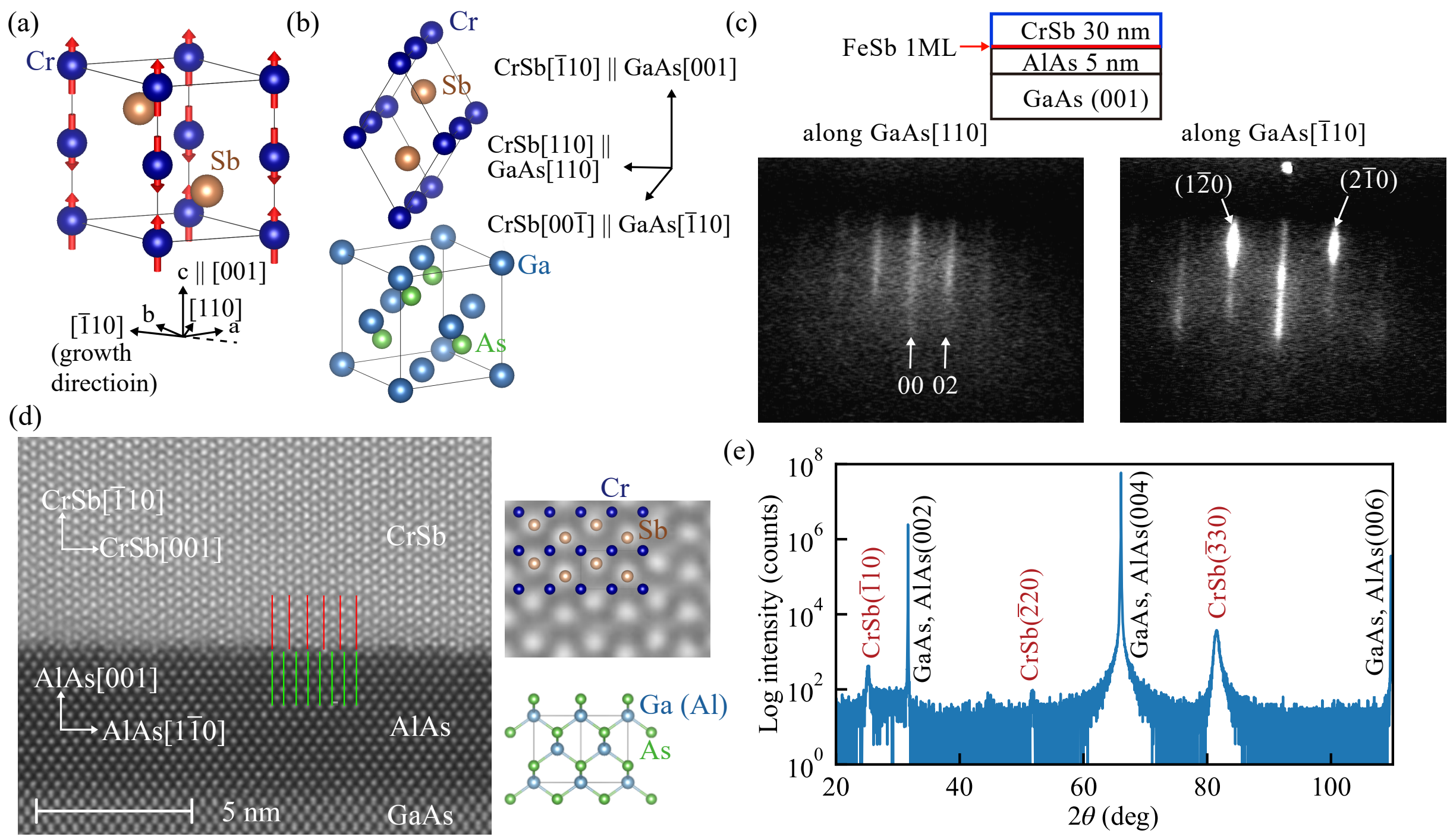

**Fig. 1.** (a) Crystal structure and magnetic configuration of CrSb drawn by VESTA [67]. The red arrows indicate the magnetic moments of Cr atoms. (b) Schematic crystal structures and epitaxial growth relationship of the CrSb film and the GaAs substrate. (c) Sample structures and RHEED patterns of the CrSb film after the growth (after post-growth annealing). (d) STEM lattice image of the sample taken along the GaAs [110] direction. Right figures are crystal structures of CrSb and GaAs (AlAs) projected along the GaAs [110] axis. Red and green vertical lines are lattice spacings near the CrSb/III-V interface. (e) XRD spectrum of the sample. The uplift near the peaks of GaAs(002) and AlAs(002) at $2\theta = 25 – 30^{\circ}$ is the fringe of the AlAs layer.

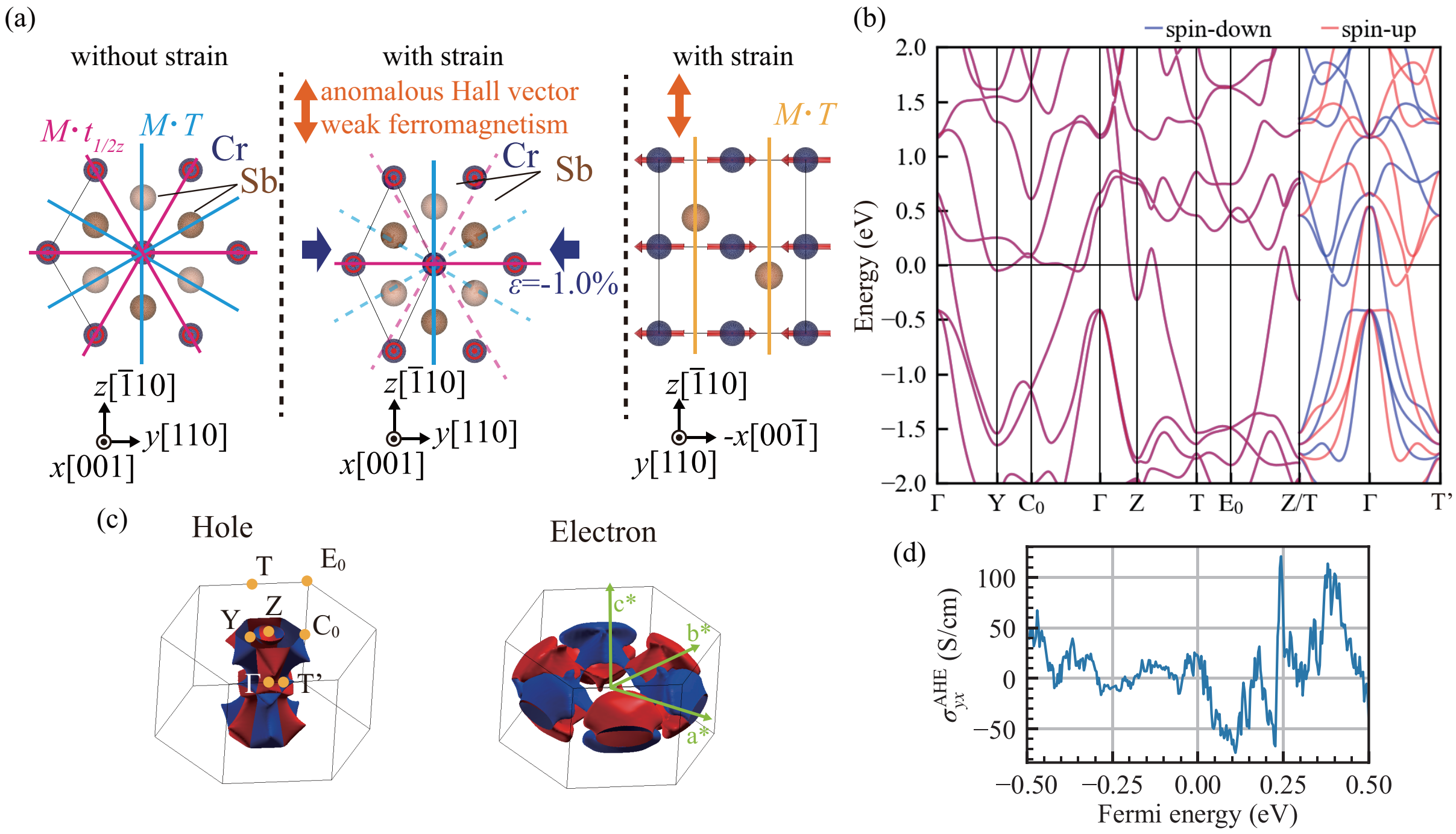

**Fig. 2.** (a) Schematic crystal structure of CrSb projected along the [001] and [110] axes, and symmetry planes of unstrained and strained CrSb (side views of the CrSb($\bar{1}10$) film). Magnetic moments are shown by red dots (⊙) indicating outward direction in the left and center panels) and red arrows (in the right panel). The symmetry planes are indicated by solid cyan, magenta, and yellow lines. *M* represents the mirror reflection operation. $t_{1/2}$ denotes the half-a-unit-cell translational operation along the CrSb [001]. *T* is the time-reversal operation. The symmetry planes broken by the strain ($\varepsilon$) applied along the [110] direction are represented by the four dashed lines. $M \cdot T$ operation reverses the direction of a *T*-odd axial vector orthogonal to the mirror plane and $M \cdot t_{1/2}$ operation reverses the direction of a *T*-odd axial vector parallel to the glide plane, and so a *T*-odd axial vector cannot have a non-zero value in any direction in unstrained CrSb [38]. The symmetry change of the strained CrSb allows the appearance of non-zero T-odd vectors, such as anomalous Hall vector and weak ferromagnetic moment, parallel to the CrSb[$\bar{1}10$] direction (orange double-headed arrows in the center and right figures). (b) Scalar-relativistic calculation of the spin-splitting band structure of CrSb. The spin-up band (red) and the spin-down band (blue) are plotted with transparency of 0.5, so the purple lines in the high-symmetry paths indicate spin degeneracy. The high symmetry positions in the first Brillouin zone (BZ) are Γ (0, 0, 0), Y (1/2, -1/2, 0), $C_0$ (0.6683, -0.3317, 0), A (0, 0, 1/2), T (1/2, -1/2, 1/2), $E_0$ (0.6683, -0.3317, 0) and T' ( -1/2, 1/2, 0) in scaled reciprocal cell vectors, respectively. (c) Spin-resolved Fermi surfaces in k-space. (d) Calculated anomalous Hall conductivity $\sigma_{yx}^{AHE}$ *vs*. Fermi energy $E_F$.

**Table. I.** Magnetic anisotropy energy of the CrSb with the lattice constants measured by XRD and the bulk crystal [34]. $E_{110}, E_{\bar{1}10}$ and $E_{001}$ indicate the total energy with the Néel vector parallel to the direction of each Miller index. The positive values indicate that the easy magnetization axis is in the [001] direction.

| | Measured lattice constants | Bulk lattice constant |
|---|---|---|
| $E_{110} - E_{001}$ (meV / unit cell) | 0.891 | 1.053 |
| $E_{\bar{1}10} - E_{001}$ (meV / unit cell) | 0.840 | 1.053 |

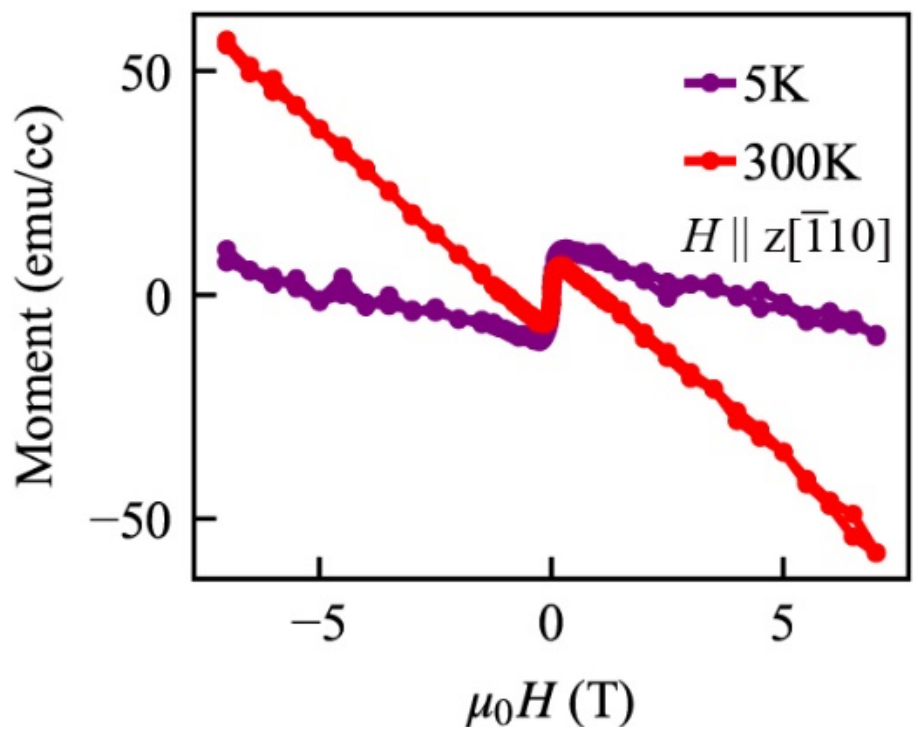

**Fig. 3.** Magnetic field dependence of the magnetization of the CrSb film measured at 5 K and 300 K. Since the diamagnetism of the GaAs substrate was much larger than the magnetization of the CrSb film, we subtracted the magnetization of the GaAs substrate measured after the removal of the CrSb film by HCl etching from the raw magnetization value, and obtained the results of Fig. 3 (see also Supplementary Fig. S5 for the detailed procedure [60,68]).

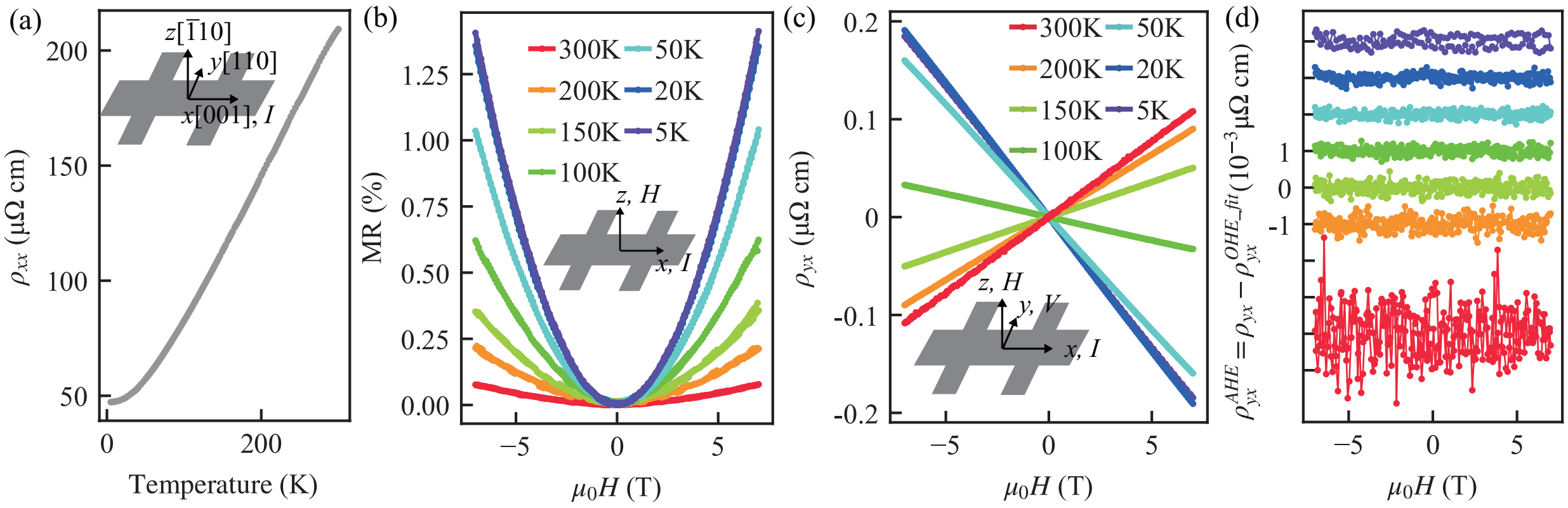

**Fig. 4.** (a) Temperature dependence of the zero-field longitudinal resistivity $\rho_{xx}$. (b), (c) Magnetic field dependence of the magnetoresistance (MR) ratio and Hall resistivity $\rho_{yx}$ measured at various temperatures when a magnetic field was applied perpendicular to the film plane ($\mu_0 H \parallel z$) and the current is along the CrSb[001] axis ($I \parallel x$). (d) The AHE component obtained by subtracting the OHE contribution, which was calculated using the carrier density and mobility obtained via a two-carrier model fit, from the original Hall resistivity. For visibility, the origin of the vertical axis is shifted for each data.

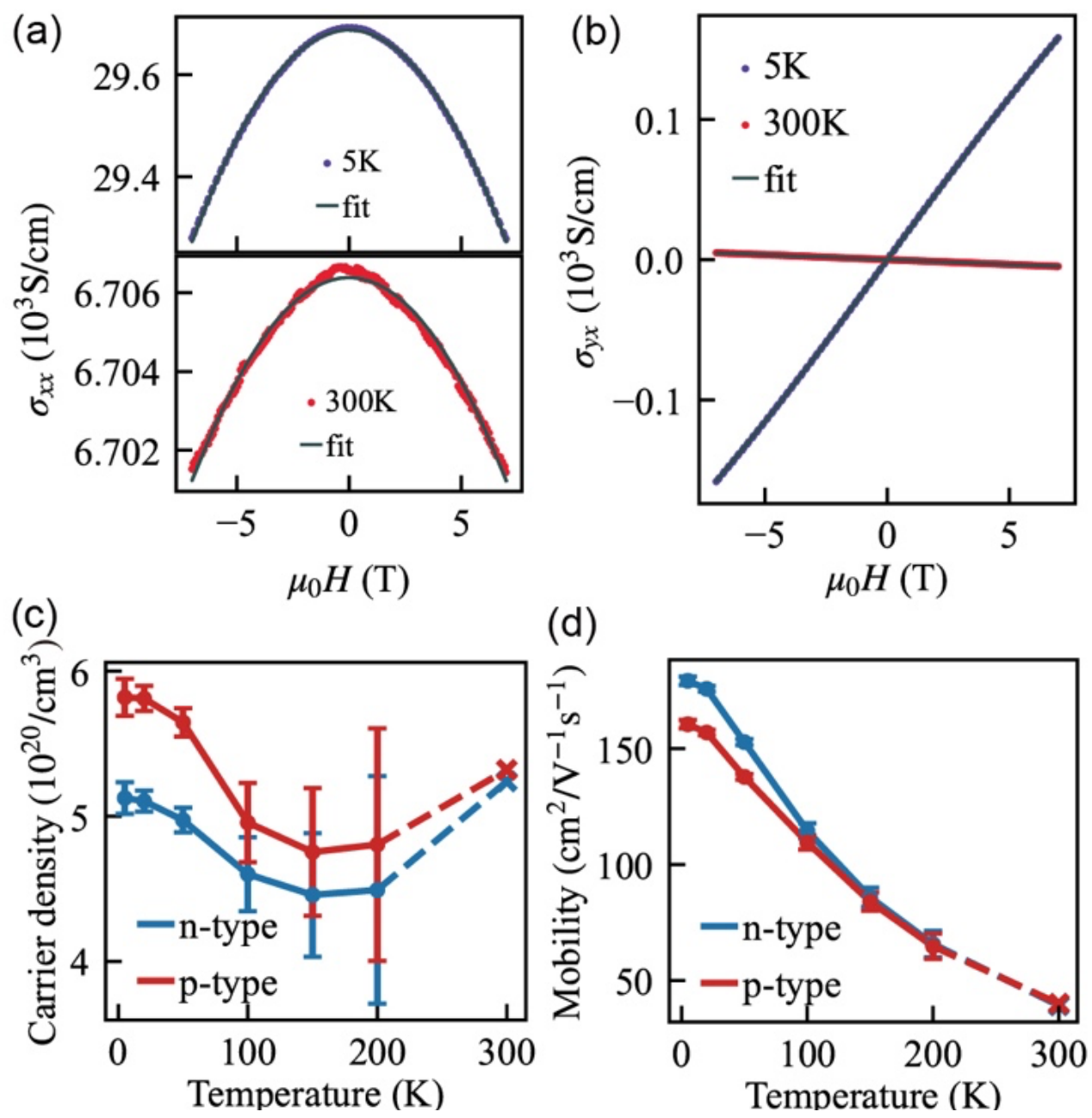

**Fig. 5.** (a), (b) Magnetic field dependences of the longitudinal conductivity $\sigma_{xx}$ and the Hall conductivity $\sigma_{yx}$ at 5 K and 300 K. The black solid lines are fitting curves based on the two-carrier model. (c), (d) Temperature dependence of the carrier (electron, hole) density and mobility extracted from the fitting of the conductivities. The standard deviation of each data points are shown by error bars. The error bar of the data point at 300 K is not plotted due to the too large standard deviation.

## Supplementary Material

# Epitaxial growth and transport properties of a metallic altermagnet CrSb on a GaAs (001) substrate

Seiji Aota[1], and Masaaki Tanaka[1,2,3,*]

[1] *Department of Electrical Engineering and Information Systems, The University of Tokyo,*
*7-3-1 Hongo, Bunkyo-ku, Tokyo 113-8656, Japan*
[2] *Center for Spintronics Research Network (CSRN), The University of Tokyo,*
*7-3-1 Hongo, Bunkyo-ku, Tokyo 113-8656, Japan*
[3] *Institute for Nano Quantum Information Electronics, , The University of Tokyo,*
*4-6-1 Komaba, Meguro-ku, Tokyo 153-0041, Japan*
[*] E-mail: masaaki@ee.t.u-tokyo.ac.jp

### Appendix
### A: Effect of 1 monolayer FeSb insertion

Another sample was grown under the same conditions as the sample in Fig. 1 except for the absence of the insertion of 1 monolayer (ML) FeSb. The XRD measurement reveals the existence of the phase of CrSb($\bar{1}11$) and CrSb($\bar{1}12$) as shown in Fig. S1.

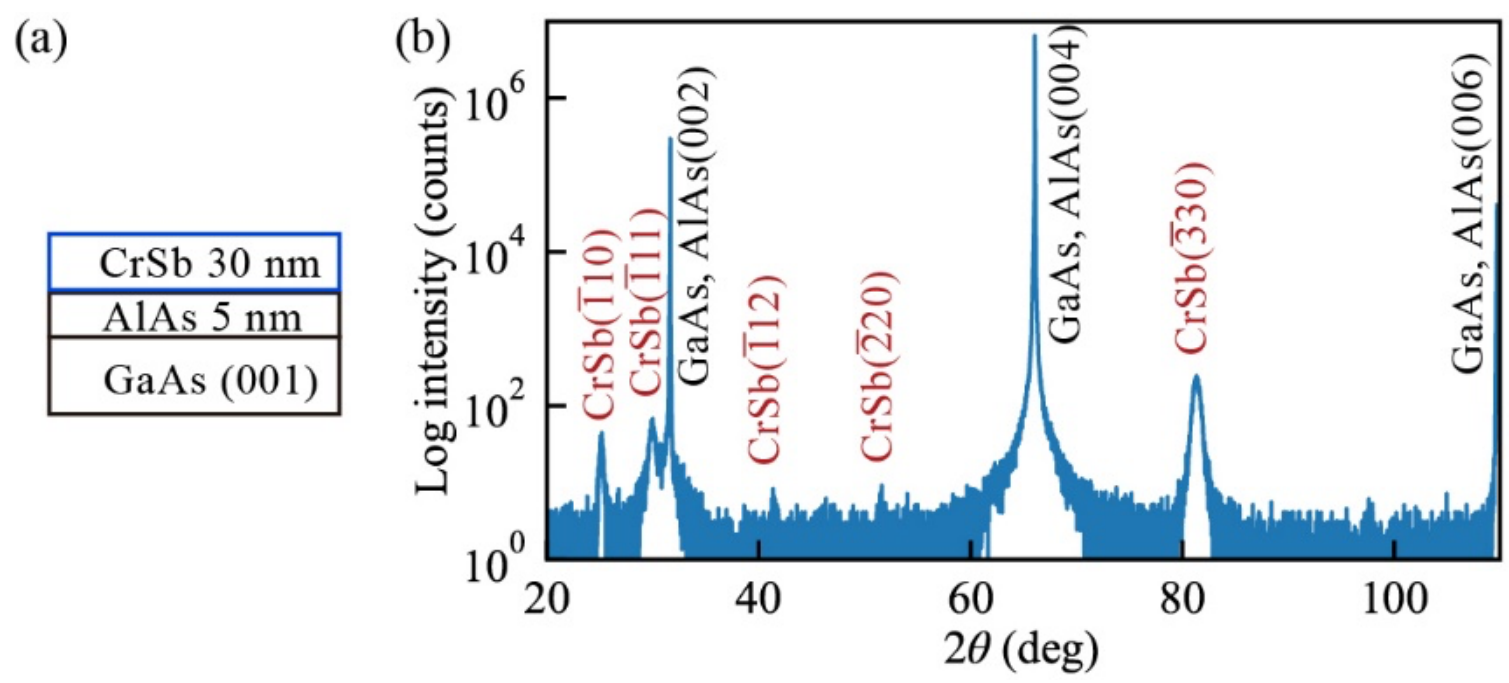

**Fig. S1.** (a) Sample structure without the insertion of 1 ML-thick FeSb but other growth conditions are the same as the sample of Fig. 1 in the main text. (b) XRD spectrum of the sample of (a).

**B: Additional structural characterization data**

We confirmed the single crystallinity and strain of the sample from STEM images of the entire film and additional XRD measurements (phi-scan and reciprocal space mapping).

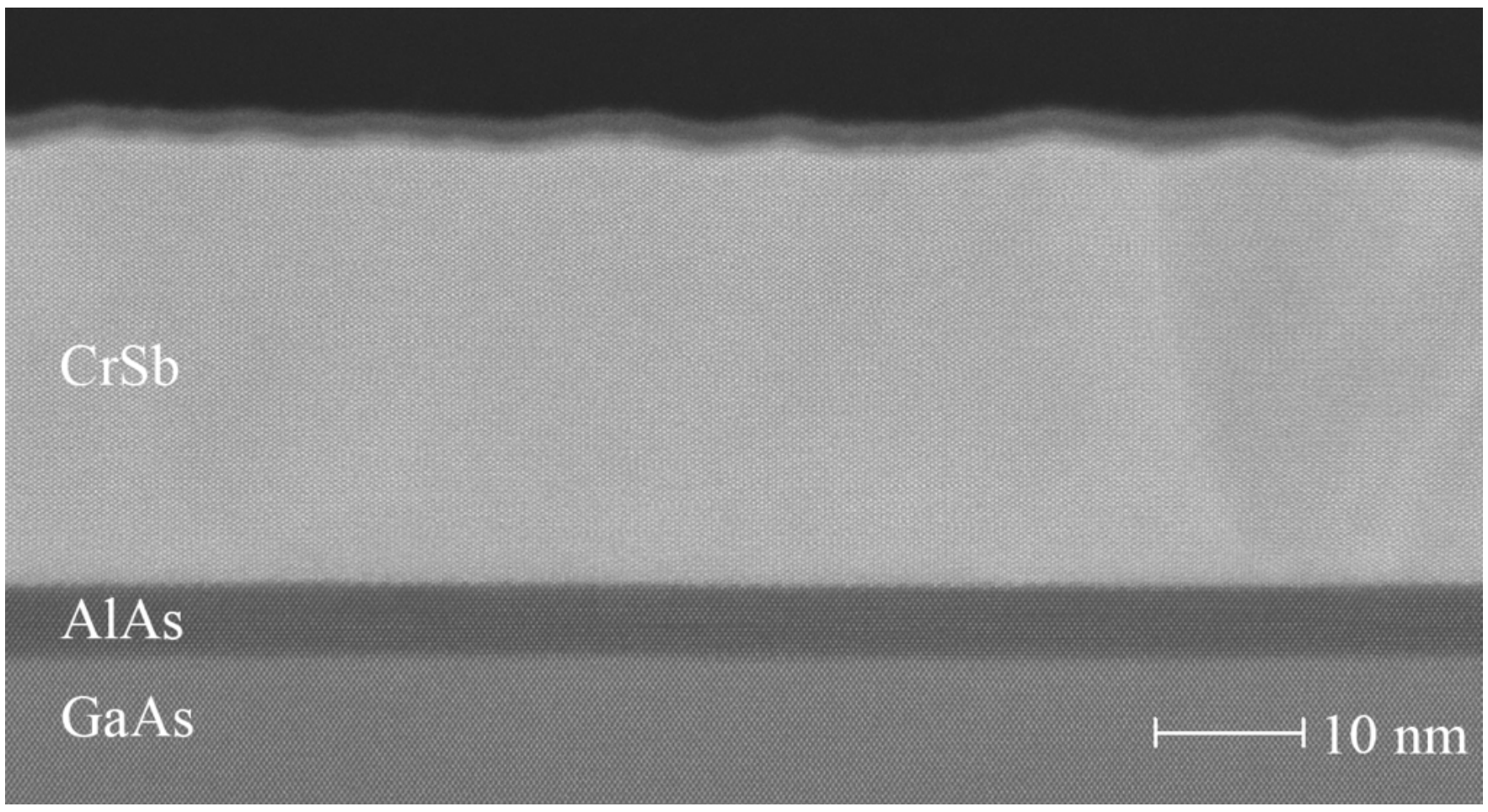

**Fig. S2.** STEM lattice image of the entire film of the 30-nm thick CrSb taken along the GaAs [110] direction.

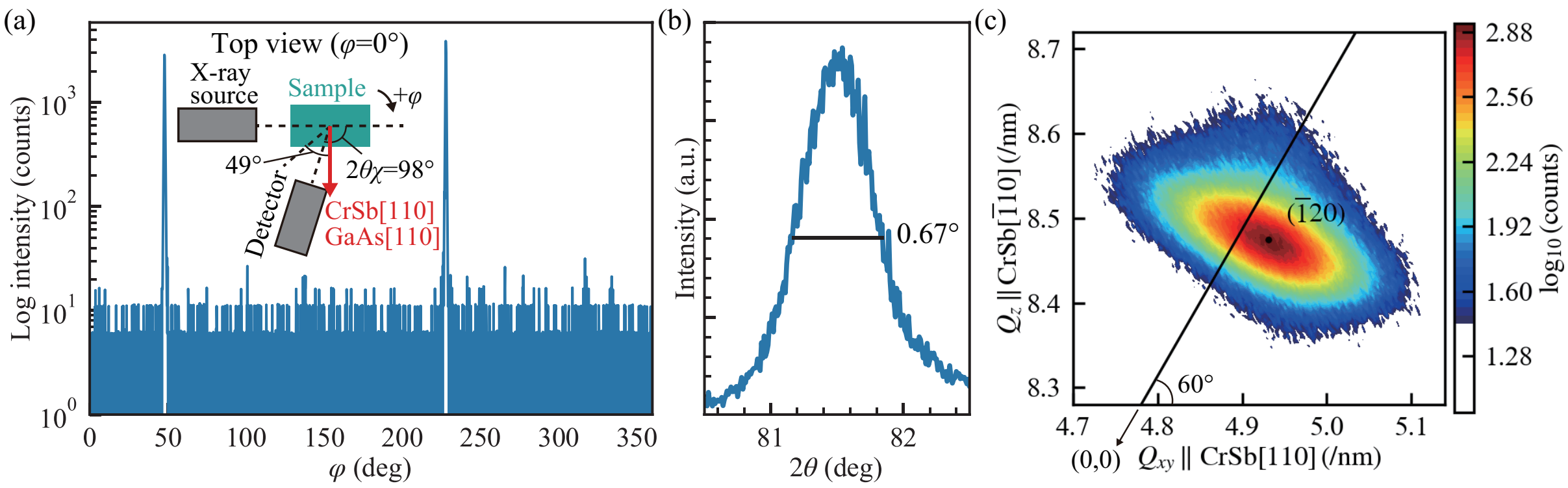

**Fig. S3.** XRD peak profiles of the sample. (a) Phi-scan patterns of the CrSb (220) peak. The inset shows the measurement setup. The $2\theta\chi$ angle was carefully determined while we avoided observing the GaAs (440) peak at $2\theta\chi = 101°$ but observed the CrSb (220) peak. (b) $\theta - 2\theta$ scan of the sample around the CrSb ($\bar{3}30$) peak. The full width at half maximum (FWHM) was estimated to be 0.67°. (c) Reciprocal space mapping around the ($\bar{1}20$) peak. The black line indicates the peak position if CrSb retained its hexagonal crystal structure.

**C: Band structure calculation of CrSb without strain**

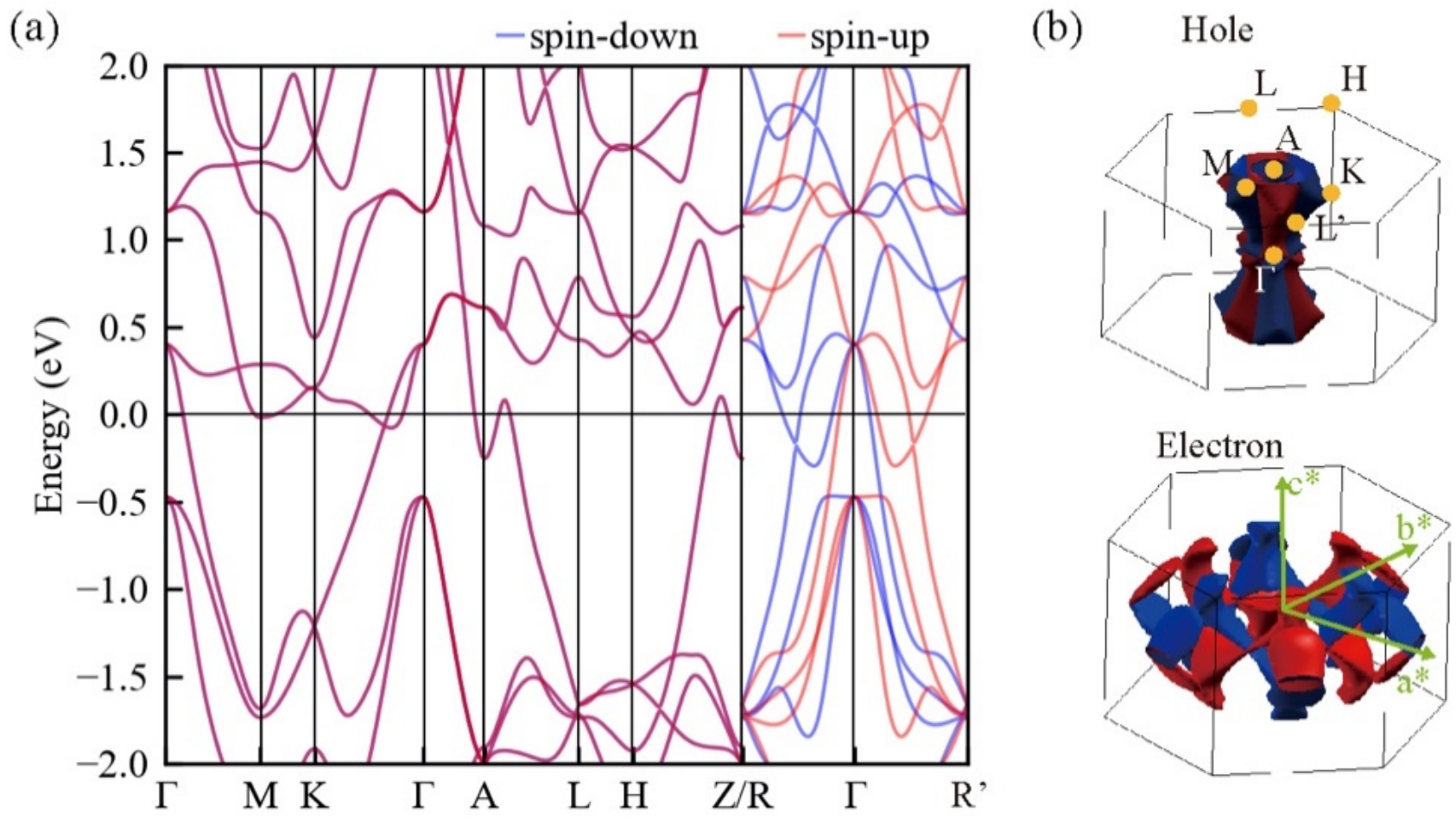

**Fig. S4.** (a) Scalar-relativistic calculation of the spin-splitting band structure of CrSb with the lattice parameters of the bulk crystal ($a$ = 0.4103 nm and $c$ = 0.5463 nm [S1]). The spin-up band (red) and the spin-down band (blue) are plotted with transparency of 0.5. The purple lines in the high-symmetry paths indicate spin degeneracy. The high symmetry positions in the first Brillouin zone (BZ) are Γ (0, 0, 0), M (1/2, -1/2, 0), K (2/3, -1/3, 0), A (0, 0, 1/2), L (1/2, -1/2, 1/2), H (2/3, -1/3, 0) and L' ( -1/2, 1/2, 0) in scaled reciprocal cell vectors, respectively. (b) The spin-resolved Fermi surfaces.

## D: Measurement method of the magnetization (*M-H*) curve of the CrSb layer

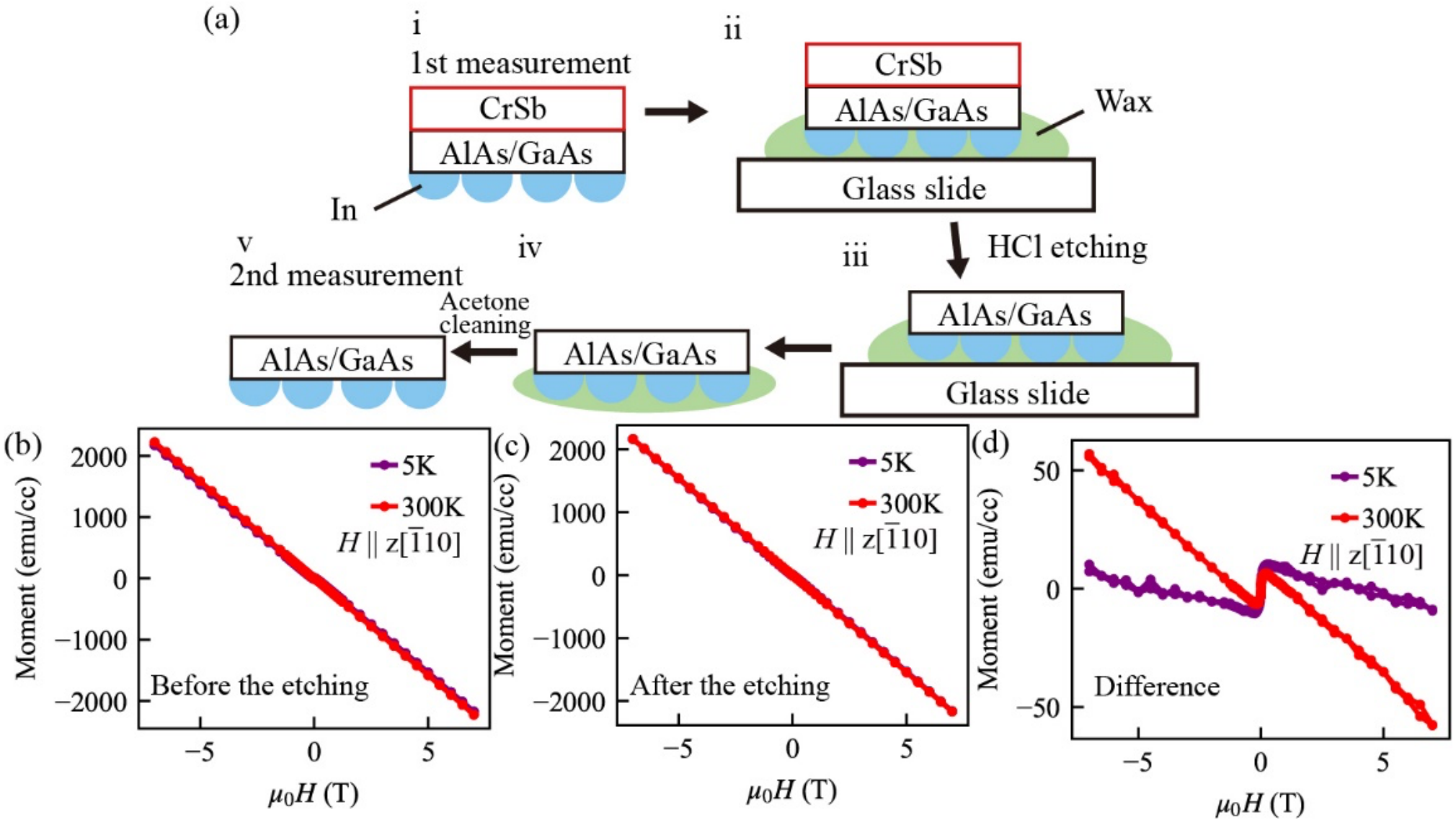

**Fig. S5.** (a) Measurement procedure to obtain the results in Fig. 3. Since the raw signal detected by the SQUID mainly came from the diamagnetism of the substrate, the CrSb layer was dissolved in a 18% HCL solution at 70 °C for 3 minutes, then the magnetization – magnetic field (*M-H*) curve was re-measured and subtract it from the raw *M-H* curve to obtain the *M-H* curve of the CrSb layer shown in Fig. 3. Removal by the dissolution of the CrSb layer was confirmed by the change in conductivity. It is known that GaAs is not drastically etched by HCL [S2], but since In was used to fix the substrate during the MBE growth, the back surface was protected with wax during the etching to prevent In from dissolving. The wax was removed with acetone solution after the etching. (b) *M-H* curve of the original sample including the substrate. (c) *M-H* curve of the sample after the removal of the CrSb layer. (d) *M-H* curve of the CrSb layer obtained after the subtraction. While positive susceptibility is expected for antiferromagnets, the subtracted signal is so small that the apparent negative susceptibility is likely attributable to measurement errors or unintended dissolution of indium.

## E: CrSb with higher annealing temperature

We investigated whether annealing at a higher temperature of 550°C would affect the AHC. The sample structure and the growth conditions other than the annealing temperature and Sb flux during the annealing were the same as the conditions of the sample described in the main text. When the sample was annealed without Sb flux, the XRD peaks of CrSb disappeared due to Sb detachment, so the new sample was annealed with an Sb flux of $1.0 \times 10^{-4}$ Pa, and subsequent measurements were performed on this sample. A small XRD peak of a secondary phase appeared at $2\theta = 45°$, the strain was decreased ($\varepsilon$ = -0.7 %), and eventually no AHC was observed either as shown in Fig. S4.

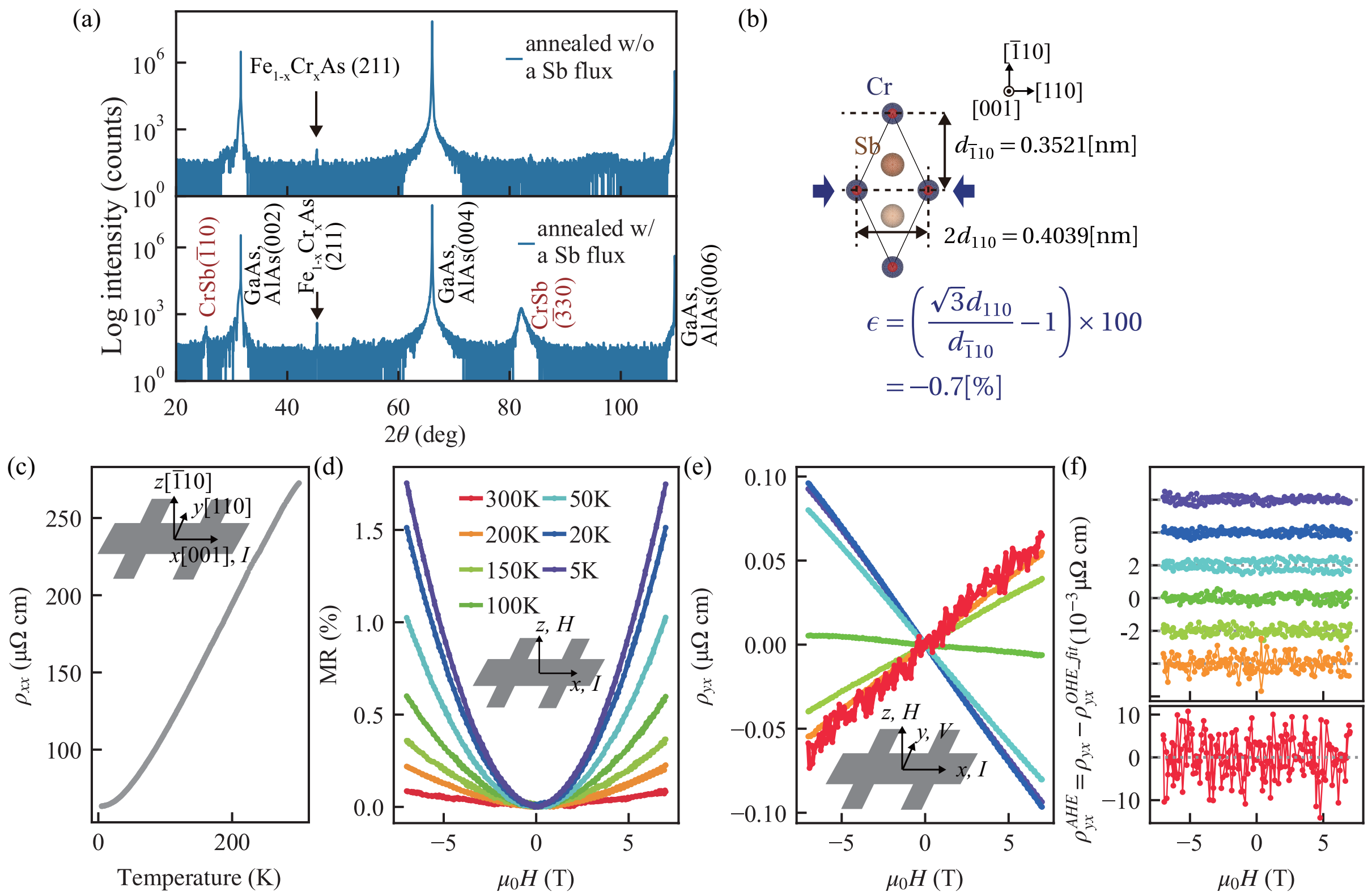

**Fig. S6.** (a) XRD spectra of the samples annealed at 550 °C for 15 minutes with and without a Sb flux of $1.0 \times 10^{-4}$ Pa. In the absence of Sb flux, the XRD peaks of CrSb disappeared and the conductivity was lost. All subsequent measurements (b) – (d) were performed on the sample annealed with a Sb flux. A small XRD peak of a secondary phase is observed at $2\theta = 45.4°$. It is attributed to $Fe_{1-x}Cr_xAs$ (211), because the observed peak position is close to $2\theta = 45.5°$ for FeAs [S3] and $2\theta$ = 43.9° for CrAs [S4] (both FeAs and CrAs have a MnP-type crystal structure) and growth of FeAs (211) on a GaAs (001) substrate was reported [S5]. (b) Schematic view of the strain and lattice constants of the grown CrSb thin film measured by RSM. (c) Temperature dependence of the zero-field longitudinal resistivity $\rho_{xx}$. The RRR is 4.30, which is almost the same value as the sample annealed at 400 °C. (d),(e) Magnetic field dependence of the magnetoresistance ratio and Hall resistivity $\rho_{yx}$ measured at various temperatures for $B \parallel z$, $I \parallel x$. (f) The AHE component obtained by subtracting the OHE contribution, which was calculated using the carrier density and mobility via a two-carrier model fit, from the original Hall resistivity. For visibility, the origin of the vertical axis is shifted for each data.

## F: Reversal of the Néel vector in altermagnets

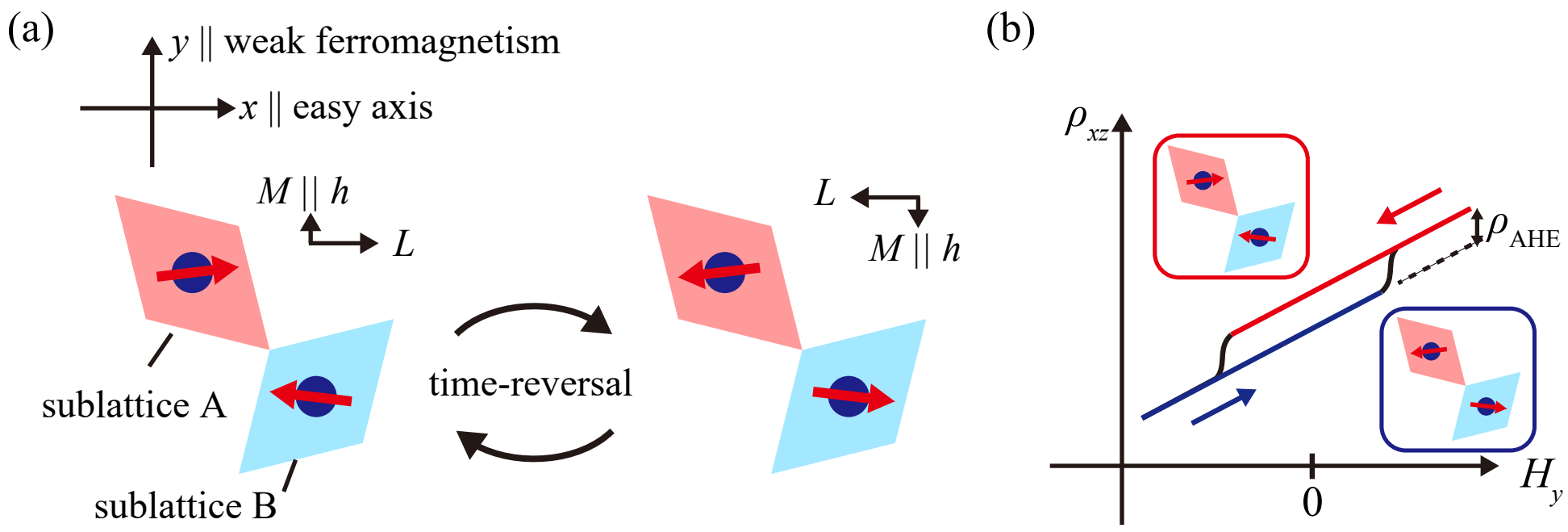

**Fig. S7.** Schematic illustrations of a reversal of the Néel vector of an altermagnet. (a) Néel vectors ($L$) parallel to the $x$-axis (easy axis of magnetization) and spin canting moment ($M$) parallel to the $y$-axis. If the spin canting direction is reversed for the opposite Néel vectors, these two states have different energy states when an external magnetic field ($H_y$) is applied parallel to the canting axis. (b) Expected result of a Hall resistivity measurement. The sign of $\rho_{AHE}$ at $H_y = 0$ (spontaneous AHE) is reversed when the Néel vector is reversed.

References

[S1] A. Kallel, H. Boller, and E. Bertaut, "Helimagnetism in MnP-type compounds: MnP, FeP, CrAs and CrAs1- xSbx mixed crystals", Journal of Physics and Chemistry of Solids 35, 1139 (1974).

[S2] M. V. Lebedev, E. Mankel, T. Mayer, and W. Jaegermann, "Wet etching of GaAs (100) in acidic and basic solutions: a synchrotron- photoemission spectroscopy study", The Journal of Physical Chemistry C 112, 18510 (2008).

[S3] K. Selte, A. Kjekshus, J. Sletten, and others, Crystal structure of FeAs, Acta Chem. Scand. **23**, 2047 (1969).

[S4] A. Jamil, H. Noguchi, and H. Munekata, Growth of Fe–As crystalline films on GaAs (100) by molecular beam epitaxy, Thin Solid Films **516**, 3015 (2008).

[S5] Z. Yu, W. Wu, Q. Hu, J. Zhao, C. Li, K. Yang, J. Cheng, J. Luo, L. Wang, and HK. Mao, Anomalous anisotropic compression behavior of superconducting CrAs under high pressure, PNAS **112**, 14766 (2015).